# Electromagnetic waves, gravitational waves and the prophets who predicted them

**Costas J. Papachristou**

Department of Physical Sciences, Hellenic Naval Academy

papachristou@snd.edu.gr

***Abstract.*** Using non-excessively-technical language and written in informal style, this article introduces the reader to the concepts of electromagnetic and gravitational waves and recounts the prediction of existence of these waves by Maxwell and Einstein, respectively. The issue of gravitational radiation is timely in view of the recent announcement of the detection of gravitational waves by the LIGO scientific team.

## *1. Introduction*

Undoubtedly, *James Clerk Maxwell* (1831-1879) and *Albert Einstein* (1879-1955) were the leading figures of Theoretical Physics in the past two centuries. Among their many achievements, Maxwell unified electricity and magnetism into a singe electromagnetic theory and predicted the existence of *electromagnetic waves*, while Einstein's Relativity changed our conception of space and time and led to modern gravitational theory, in the context of which *gravitational waves* were predicted to exist.

Unfortunately, Maxwell didn't live long enough to see the experimental confirmation of existence of electromagnetic waves. As for Einstein, in a sense he was "luckier" since it would be biologically impossible for him, anyway, to be present in the official announcement of the detection of gravitational waves one whole century after he had predicted them to exist!

In this article we will try to explain the nature of electromagnetic and gravitational waves and examine how they are produced. For further and deeper study of the subject the reader is referred to the sources cited at the end.

## *2. Maxwell and the first unification theory for interactions*

As is well known, every electric charge (regardless of its motion) produces an electric field and is itself subject to a force inside an existing electric field. Also, every *moving* charge produces a magnetic field and experiences a force inside such a field.

We often tend to view *electricity* and *magnetism* as two separate natural phenomena. Indeed, in a hypothetical world where all electric and magnetic fields remained unchanged with time, there would be no way of knowing that electric and magnetic phenomena are interrelated and mutually dependent. At the theoretical level, the set of four *Maxwell equations* [1,2] would break into two *independent* pairs, one for the electric and one for the magnetic field.

In 1831, in a series of experiments, *Michael Faraday* discovered something interesting: Whenever a magnetic field changes with time, an electric field emerges! Although there were no experimental indications at the time, Maxwell assumed that the converse is also true; that is, a magnetic field is present whenever the electric field varies with time. Thus, given this mutual dependence between the electric and the magnetic field, electric and magnetic phenomena should not be treated as separate.

Historically, this was the first *unification theory* of apparently different interactions – electric and magnetic – into a single *electromagnetic interaction*. In the twentieth century the unification scheme would be enhanced by incorporating the weak and the strong interaction – and by making a heroic, albeit frustrating, effort to include gravity as well...

### *3. Electromagnetic waves*

With his mathematical genius, Maxwell "codified" the laws of electromagnetism with a system of four equations (expressed in differential or, equivalently, in integral form) that describe the behavior of the electromagnetic field in space and time [1,2]. One consequence of these equations is the conclusion that the electromagnetic field must exhibit wavelike properties. That is, a change ("disturbance") of the field at some point of space is not felt instantaneously at other points but propagates as an *electromagnetic wave* traveling at the speed of light. In particular, light itself is just a special type of electromagnetic wave having the additional property of being sensed by our eyes.

The importance of electromagnetic waves for our lives cannot be overemphasized! Through them we receive light and warmth (and, unfortunately, some harmful radiation as well) from the Sun, we enjoy stereo music on the radio, we watch football matches on TV, we communicate by using our cell phones... But, how are these waves produced in the first place?

First, some terminology: The propagation of energy by means of electromagnetic waves is called *electromagnetic radiation*. (Henceforth we will write *"e/m wave"* and *"e/m radiation"*, for short.) Thus, a physical system that emits energy in the form of e/m waves is said to *emit e/m radiation* or, simply, to *radiate*. Such systems include atoms, molecules, nuclei, hot bodies, radio-station antennas, etc.

By a careful examination of the Maxwell equations it follows that the e/m radiation is produced in basically two ways: (*a*) by *accelerated* electric charges and (*b*) by time-varying electric currents [1,2]. In particular, a non-accelerating charge (one that moves on a straight line with constant speed) does *not* radiate. I often explain this to my students (before writing any equations) by using the following parable:

On a hot summer day you go to the store and buy an ice cream. You decide to eat it on the road before it melts. You take a carefree walk on a straight path, with steady step (thus, with *constant velocity*), without noticing a swarm of bees following you (or, rather, your ice cream)! When you suddenly notice them, you *accelerate* your motion in order to escape from them (you either move faster in the same direction or just change your direction of motion). Scared by this move of yours, some of the bees leave the swarm and fly away, never to come back...

What is the meaning of all this? The "ice cream" is an electric charge initially moving with constant velocity and carrying with it the total energy of its e/m field (the "swarm of bees"). This is just a transfer of a constant amount of energy in the direction of motion of the charge. When the charge accelerates, a part of this energy (the "bees" that fly away) is detached, in a sense, and travels to infinity at the speed of light in the form of an e/m wave. And, the higher the acceleration of the charge, the greater the energy radiated per unit time.

### *4. Einstein and Relativity*

In empty space, the speed of light (denoted *c*) is approximately 300,000 kilometers per second. Now, speeds (and, more generally, velocities) are determined relative to some *frame of reference*. For example, when a passenger walks along the corridor of a moving bus, her speed as measured by a seated passenger is different from that which would be recorded by someone standing on the sidewalk. The bus and the sidewalk define two different frames of reference relative to which the velocity of the walking passenger is determined.

So, relative to which frame of reference does the speed of light have the familiar value *c*? Based on perceptions at his time, Maxwell assumed that the speed of propagation of e/m radiation takes on the "correct" value *c* in a privileged frame of reference that is at rest relative to the *ether*, a hypothetical substance with almost metaphysical properties that was believed to occupy the whole of space. It is in this frame only that the Maxwell equations would assume their proper form. Thus, any observer moving relative to the ether should measure a speed of light different from *c* and should also conclude that the electromagnetic phenomena are not correctly described by Maxwell's equations.

However, every attempt to experimentally verify the dependence of the speed of light on the state of motion of the observer failed. Then, in a historic article of 1905, Einstein proposed that the speed of light in vacuum has the *same* value *c* for *all* observers, regardless of their state of motion. Moreover, the laws of Physics – and, in particular, the Maxwell equations of electrodynamics – should assume the same form in all frames of reference. (Technically speaking, the above principles are valid for a special class of *inertial observers* associated with *inertial frames of reference*.) These principles form the basis of the *Special Theory of Relativity*.

In classical (Newtonian) mechanics, time has absolute meaning, common to all observers. Thus, according to this theory, if a pulse of light is emitted from one point of space toward another, different observers will agree with one another with regard to the time it took for light to make the journey, although they will possibly disagree on the distance traveled (each observer will measure this distance *relative to himself*).

In Relativity, however, these observers must agree with one another regarding the speed $c$ of light. Given that they will disagree, in general, about the length of the route, they must now also disagree with regard to the time taken for light to make the trip. Thus, Relativity puts an end to the idea of absolute time. Time intervals as well as spatial distances are directly dependent on the motion of the observer and are devoid of absolute meaning.

In addition to this, the constancy of the speed of light imposes a sort of mathematical interweaving between space and time coordinates of an event, such that the distinction between space and time is also not absolute but depends on the motion of the observer. Thus, in place of the separate terms *"space"* and *"time"*, in Relativity one speaks of *spacetime*.

Special Relativity did not change the classical form of the Maxwell equations; however, it dramatically revised Newtonian Mechanics, which was now seen to be valid as an approximation in the limit of "small" speeds (in comparison, that is, to the huge value of $c$). Among other things, Relativity revealed a remarkable relationship between mass and energy ($E=mc^2$), which has no classical analog. As we know, an early "experimental" verification of this relation costed countless human lives in Hiroshima and Nagasaki at the end of WW2.

### *5. Gravity is... geometry!*

The spacetime of Relativity is four-dimensional, with three dimensions corresponding to space and one to time. Let us consider, for simplicity, just two dimensions, one for space and one for time. In Special Relativity the geometry of such a two-dimensional spacetime would *look like* that of an infinite plane surface (although the mathematical recipe for evaluating distances would be somewhat different). With regard to intrinsic geometrical properties, such a *flat* surface has fundamental differences from curved surfaces such as, e.g., the surface of a sphere.

In 1915, Einstein proposed his *General Theory of Relativity*, which was based on a very original idea: What we perceive as *gravity* is not, in reality, a force (like, e.g., electric or magnetic forces) but is a manifestation of a *geometric deformation of flat space* (technically speaking, of flat *spacetime*) caused by the presence of matter [3]. Thus, for example, the observed motion of the Earth around the Sun is not – as Newton would assert – due to the gravitational force exerted on the Earth by the Sun but is due to the *curvature of space* caused by the mass of the Sun itself. Space has no longer geometrical properties similar to these of a flat surface but rather similar to those of the curved surface of a sphere.

So, in General Relativity the gravitational field is not treated as a force field but rather as a field of deformations ("ripples") in the fabric of spacetime. And, locally, these deformations are greater the greater the mass that causes them.

## *6. Gravitational waves*

What happens when the gravity-related ripples at some region of space change in time due to a redistribution of matter in that region? Let us recall the situation in electromagnetism: Every redistribution of the sources of the e/m field (charges and/or currents) in a region of space causes a disturbance of the e/m field in that region, which disturbance propagates in space as an e/m wave at the speed of light, *c*. In particular, an *accelerated* electric charge emits energy in the form of e/m radiation. The emitted e/m wave thus takes away a part of the charge's total energy.

Now, as Einstein showed in 1916, a consequence of the equations of General Relativity is that any redistribution of matter in a region of space causes a disturbance of local geometry (in classical terms, of the gravitational field), which (disturbance) propagates in space as a *gravitational wave* [3–6] traveling at speed *c*. Moreover, an *accelerated* body loses part of its energy, as it becomes the source of *gravitational radiation*.

The problem is that, whereas even an atomic system may emit detectable e/m radiation (e.g., visible light), the production of detectable gravitational radiation requires enormous masses with very high accelerations. Such physical conditions do exist in the Universe (rotating pairs of neutron stars or black holes, stellar collisions and explosions, etc.) and the liberated gravitational energy is indeed huge. These phenomena, however, (fortunately!) occur so far from us that, by the time they reach the Earth, the emitted gravitational waves will be millions of times weaker. Thus, an exceptionally sensitive device is needed in order to detect these waves.

## *7. Why are they useful?*

Until recently, all information we obtained about the Universe was based on observations via e/m radiation (visible light, radio waves, microwaves, X-rays, etc.). Gravitational waves may now provide information that would otherwise be impossible to get. For example, a collision of black holes does not produce an appreciable amount of e/m radiation while it does produce enormous gravitational radiation. Thus, with the aid of gravitational waves we will be able to study such catastrophic cosmic phenomena.

Also, in contrast to e/m radiation, which interacts strongly with matter – thus is subject to absorption and distortion as it crosses distances of millions of light-years in the Universe – gravitational waves can travel huge distances practically unchanged (they only weaken in magnitude as they spread in space while moving away from the source). Thus, the information these latter waves provide is much more faithful compared to that furnished by e/m waves.

Finally, gravitational waves are expected to provide answers to some important cosmological questions regarding the early stages of evolution of the Universe. Traditional astronomy is not in a position to answer such questions, as the Universe was initially opaque to e/m radiation and thus no information of electromagnetic origin may reach us from that cosmic period.

## *8. Epilog: Why all this excitement lately?*

Even though Einstein had predicted the existence of gravitational waves as early as in 1916, an *indirect* astronomical confirmation of their existence was obtained much later, in the 1970s. However, there had never been a *direct* detection of such waves on Earth.

On February 11, 2016, the scientific team of *LIGO* (*Laser Interferometer Gravitational-wave Observatory*) [4] announced that they had detected gravitational waves on September 14, 2015. These waves were produced by the merging of two black holes (which initially formed a rotating pair) at a distance of *1.3 billion light-years* from Earth [7]. It was the ultimate confirmation of Einstein's General Relativity!

The LIGO "observatory" consists of two identical detectors (*laser interferometers* [4]) located in the USA, with a distance of about 3,000 kilometers between them. *LIGO Hanford* is in southeastern Washington State; *LIGO Livingston* is in Louisiana.

Certainly this is not the end of the story. It is only a small but decisive first step in man's deep cosmic experience, the beginning of an ambitious endeavor to explore the *"final frontier"* of space – *"to boldly go where no man has gone before"* !

## *Notes and References*

[1] Manuscript: https://arxiv.org/abs/1711.09969

[6] Video: *Gravitational Waves Explained* (https://youtu.be/4GbWfNHtHRg).

[7] Because of its rotation, the system of the two *accelerating* black holes continuously lost energy as it emitted gravitational waves. This had the effect of diminishing the distance between the two objects, which fact made the system spin faster and faster and emit more and more gravitational radiation in the process, which again forced the two bodies to get closer and closer, and so forth. At the final stage of the process, the two objects collided and merged, emitting a huge quantity of gravitational energy within a very short time.